\documentclass[twocolumn,floatfix,aps]{revtex4}
\usepackage{graphicx}
\usepackage{amsmath}
\usepackage{amssymb}
\usepackage{bm}
\usepackage{hyperref}

\begin{document}

\title{Chirality inversion of Majorana edge modes in a Fu-Kane heterostructure}
\author{A. Don\'{i}s Vela}
\affiliation{Instituut-Lorentz, Universiteit Leiden, P.O. Box 9506, 2300 RA Leiden, The Netherlands}
\author{G. Lemut}
\affiliation{Instituut-Lorentz, Universiteit Leiden, P.O. Box 9506, 2300 RA Leiden, The Netherlands}
\author{M. J. Pacholski}
\affiliation{Instituut-Lorentz, Universiteit Leiden, P.O. Box 9506, 2300 RA Leiden, The Netherlands}
\author{C. W. J. Beenakker}
\affiliation{Instituut-Lorentz, Universiteit Leiden, P.O. Box 9506, 2300 RA Leiden, The Netherlands}
\date{May 2021}
\begin{abstract}
Fu and Kane have discovered that a topological insulator with induced \textit{s}-wave superconductivity (gap $\Delta_0$, Fermi velocity $v_{\rm F}$, Fermi energy $\mu$) supports chiral Majorana modes propagating on the surface along the edge with a magnetic insulator. We show that the direction of motion of the Majorana fermions can be inverted by the counterflow of supercurrent, when the Cooper pair momentum along the boundary exceeds $\Delta_0^2/\mu v_{\rm F}$. The chirality inversion is signaled by a doubling of the thermal conductance of a channel parallel to the supercurrent. Moreover, the inverted edge can transport a nonzero electrical current, carried by a Dirac mode that appears when the Majorana mode switches chirality. The chirality inversion is a unique signature of Majorana fermions in a spinful topological superconductor: it does not exist for spinless chiral \textit{p}-wave pairing.
\end{abstract}
\maketitle

\section{Introduction}

The chiral edge modes of the quantum Hall effect in a semiconductor have a superconducting analogue \cite{Kal16}: A two-dimensional (2D) superconductor with broken time-reversal symmetry and broken spin-rotation symmetry can enter a phase in which the gapped interior supports gapless edge excitations. This is called a topological superconductor, because the number of edge modes is set by a topological invariant \cite{Qi11,Bee16,Sat17}. Each edge mode contributes a quantized unit of thermal conductance, producing the thermal quantum Hall effect \cite{Rea00}. The edge modes are referred to as Majorana modes, since the quasiparticle excitations at the Fermi level are their own antiparticle --- being equal-weight superpositions of electrons and holes. 

Chiral edge modes have not yet been conclusively observed in a superconductor \cite{Mac17,Kay20}, due in part to the complexity of heat transport measurements at low temperatures. In this work we propose an \textit{electrical} signature of a chiral edge mode, triggered by the chirality inversion when a supercurrent flows along the boundary.

Our study was motivated by the recent experimental observation of the Doppler effect from a superflow in a topological superconductor \cite{Zhu20}. The 2D electron gas of massless Dirac fermions on the surface of the topological insulator Bi$_2$Te$_3$ is proximitized by the superconductor NbSe$_2$, so that a gap $\Delta_0$ opens up at the Fermi level $\mu$. An in-plane magnetic field $B$ induces a screening supercurrent over a London penetration depth $\lambda_{\rm L}$, which boosts the Cooper pair momentum by an amount $K \simeq eB\lambda_{\rm L}$, in-plane and perpendicular to $B$. The Doppler effect \cite{Tinkham,Vol07} shifts the quasiparticle energy by $\delta E=v_{\rm F}K$, closing the gap when $K$ exceeds $K^\ast=\Delta_0/v_{\rm F}$ \cite{Yua18,Pap21,Pac21}.

\begin{figure}[tb]
\centerline{\includegraphics[width=1\linewidth]{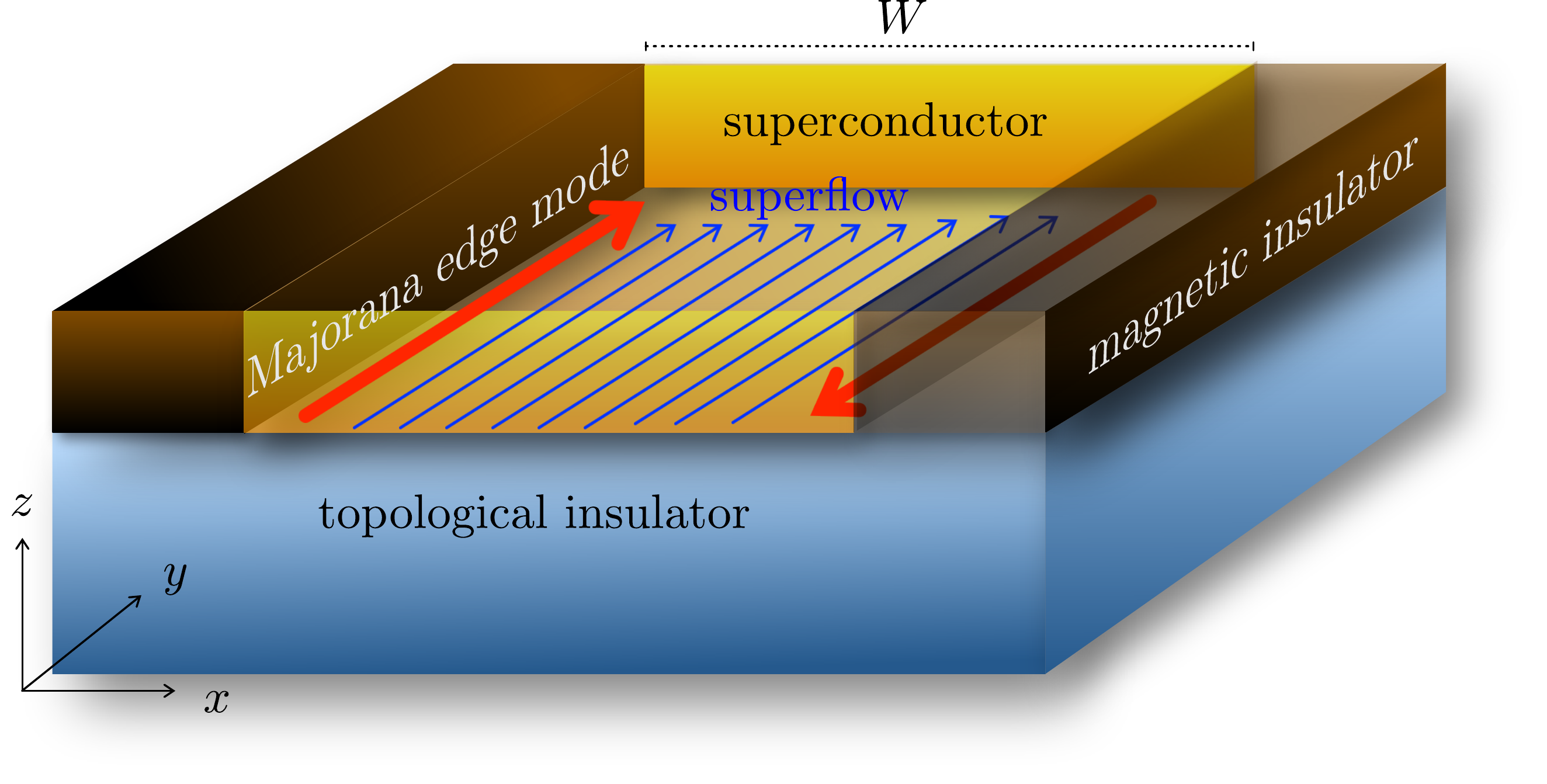}}
\caption{Schematic of the Fu-Kane heterostructure \cite{Fu08}, a topological insulator with induced \textit{s}-wave superconductivity (gap $\Delta_0$, Fermi velocity $v_{\rm F}$, Fermi energy $\mu$). The surface electrons are confined to a channel (width $W$) by a pair of magnetic insulators with perpendicular magnetization. Counterpropagating Majorana edge modes are indicated by red arrows. The blue arrows indicate the superflow (Cooper pair momentum ${K}$). The Doppler effect boosts the velocity of the Majorana mode on the left edge, while the right edge mode slows down and inverts its direction of motion when $K>\Delta_0^2/\mu v_{\rm F}$. At that chirality inversion a Dirac mode appears on the right edge, moving oppositely to the superflow.
}
\label{fig_layout}
\end{figure}

The ingredient we add to the system of Ref.\ \onlinecite{Zhu20} is the confinement produced by a magnetic insulator (EuS) with magnetization perpendicular to the surface layer (see Fig.\ \ref{fig_layout}). This is the Fu-Kane proposal \cite{Fu08} for chiral Majorana modes. Our key finding is that the superflow inverts the chirality of a Majorana mode moving in the opposite direction once $K$ exceeds $K_c=K^\ast\Delta_0/\mu$ --- so well before the gap closing transition for $\Delta_0\ll\mu$. This chirality inversion can be detected in a transport experiment, both in thermal and in electrical conduction.

\section{Chirality inversion}

We base our analysis on the 2D Dirac-Bogoliubov-de Gennes Hamiltonian of a topological insulator surface (Fermi energy $\mu=v_{\rm F}k_{\rm F}$, $\hbar\equiv 1$) with induced \textit{s}-wave superconductivity at Cooper pair momentum $\bm{K}$,
\begin{equation}
{\cal H}=\bigl(v_{\rm F}\bm{k}\cdot\bm{\sigma}-\mu\sigma_0 \bigr) \tau_z+\bigl(v_{\rm F}\bm{K}\cdot\bm{\sigma}+M\sigma_z\bigr) \tau_0+\Delta_0\sigma_0 \tau_x.\label{Hfourband}
\end{equation}
The vectors $\bm{k}$, $\bm{\sigma}$, $\bm{K}$ have only $x$ and $y$ components, in the plane of the surface. The magnetization $M$ points in the $z$-direction. The $\sigma$ and $\tau$ Pauli matrices act, respectively, on spin and electron-hole degrees of freedom \cite{Rashba}.

We confine the electrons to a strip of width $W$ parallel to the $y$-axis, by setting $M=0$ for $|x|<W/2$ and $M\rightarrow+\infty$ for $|x|>W/2$. Integrating
\begin{equation}
-iv_{\rm F}\sigma_x \tau_z\partial_x\psi=-M\sigma_z \tau_0\psi\Rightarrow v_{\rm F}\partial_x\psi=-M\sigma_y \tau_z\psi
\end{equation}
from $x=\pm W/2$ to $\pm\infty$, and demanding a decaying wave function, we obtain the boundary condition
\begin{equation}
\psi(x,y)=\pm \sigma_y \tau_z\psi(x,y)\;\; \text{at}\;\;x=\pm W/2.\label{boundcond}
\end{equation}
The spinor structure of the wave function at the boundaries is therefore a superposition of
\begin{equation}
| u_1\rangle=\begin{pmatrix}
i\\
1
\end{pmatrix}\otimes
\begin{pmatrix}
1\\
0
\end{pmatrix},\;\;| u_2\rangle=\begin{pmatrix}
-i\\
1
\end{pmatrix}\otimes
\begin{pmatrix}
0\\
1
\end{pmatrix}\label{u1u2def}
\end{equation}
at $x=-W/2$ and a superposition of $\tau_x|u_1\rangle$, $\tau_x|u_2\rangle$ at $x=W/2$.

We seek the wave function profile $\psi(x,y)=e^{ik_y y}\psi(x)$ at energy $E$ and wave vector $k_y$ parallel to the boundary. The superflow momentum $\bm{K}=(0,K)$ is oriented along the boundary.  Integration of the Schr\"{o}dinger equation ${\cal H}\psi=E\psi$ gives $\psi'(x)=\Omega\psi(x)\Rightarrow \psi(W/2)=e^{W\Omega}\psi(-W/2)$ with
\begin{align}
\Omega={}&i(E/v_{\rm F})\sigma_x \tau_z+i(\mu/v_{\rm F})\sigma_x \tau_0-iK_x\sigma_0 \tau_z+K\sigma_z \tau_z\nonumber\\
&+k_y\sigma_z \tau_0+(\Delta_0/v_{\rm F})\sigma_x \tau_y.\label{Omegadef}
\end{align}

The boundary condition \eqref{boundcond} dictates that $\psi(-W/2)$ is a superposition of the states $| u_1\rangle$, $| u_2\rangle$, while $\psi(W/2)$ is orthogonal to these two states. This gives the determinantal equation
\begin{equation}
{\rm det}\,\Xi=0,\;\;\Xi_{nm}=\langle  u_n|e^{W\Omega}| u_m\rangle,\label{detXi}
\end{equation}
from which we determine the spectrum $E(k_y)$. In the limit $W\rightarrow\infty$ of uncoupled edges we find near $k_y=0$ the Majorana edge mode dispersion \cite{appendices}
\begin{align}
E_\pm&=\pm v_{\rm F}k_y\frac{\Delta_0^2+v_{\rm F}^2K^2\pm v_{\rm F}K\mu}{\Delta_0^2\mp v_{\rm F}K\mu+\mu^2}\nonumber\\
&\rightarrow v_{\rm F}k_y(K/k_{\rm F}\pm\Delta_0^2/\mu^2) \;\;\text{for}\;\;\mu\gg\Delta_0.\label{Ekylargemu}
\end{align}
The $\pm$ sign distinguishes the modes on opposite edges. These are Majorana modes, because they are nondegenerate and transform into themselves when charge conjugation maps $E\mapsto -E$ and $k_y\mapsto -k_y$.

\begin{figure}[tb]
\centerline{\includegraphics[width=1\linewidth]{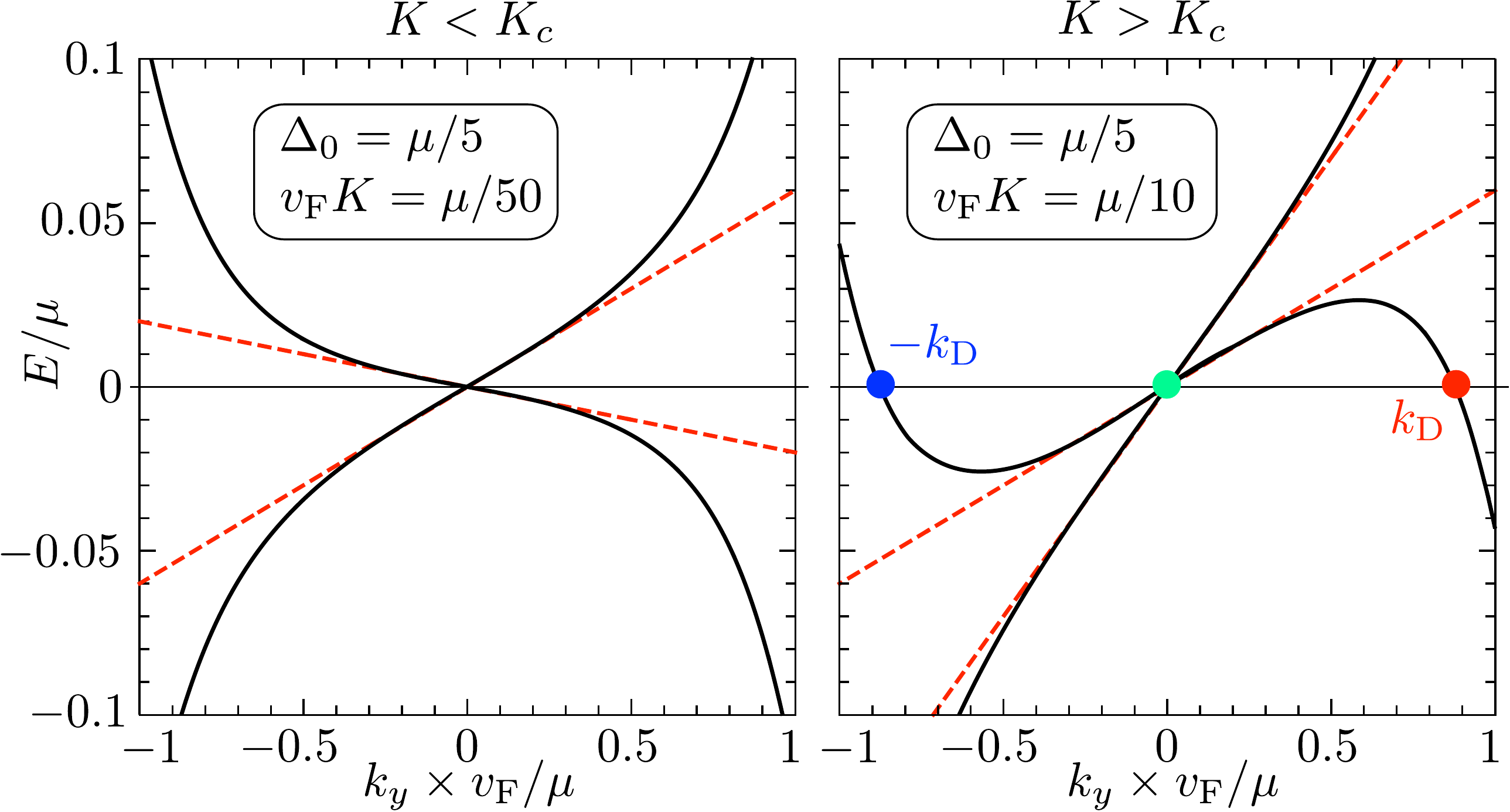}}
\caption{Dispersion relation of the edge modes in the non-inverted regime ($K<K_c$) and in the inverted regime ($K>K_c$). The solid curves are calculated numerically from Eq.\ \eqref{detXi} for channel width $W=100\,v_{\rm F}/\mu$. The dashed lines are the large-$\mu$, large-$W$ asymptotes \eqref{Ekylargemu}. The colored dots in the inverted regime indicate the charge-neutral Majorana mode (at $k_y=0$) and the electron-like and hole-like Dirac modes (at $k_y=\pm k_{\rm D}$).
}
\label{fig_dispersion1}
\end{figure}

The group velocity of an edge mode equals $dE/dk_y$, and hence we conclude from Eq.\ \eqref{Ekylargemu} that a chirality inversion appears with increasing $K$, such that for $K>K_c$ both Majorana edge modes propagate in the same direction. This is illustrated in Fig.\ \ref{fig_dispersion1}. The critical $K_c$ equals
\begin{equation}
K_c=\frac{2\Delta_0^2/v_{\rm F}}{\sqrt{4\Delta_0^2+\mu^2}+\mu}\rightarrow\frac{\Delta_0^2}{v_{\rm F}\mu}\;\;\text{for}\;\;\mu\gg\Delta_0.\label{Kcexact}
\end{equation}
Since the gap in the bulk spectrum does not close until $K=K^\ast=\Delta_0/v_{\rm F}$ the bulk remains gapped in the inverted regime --- only the edge modes propagate at the Fermi energy ($E=0$).

\begin{figure}[tb]
\centerline{\includegraphics[width=0.8\linewidth]{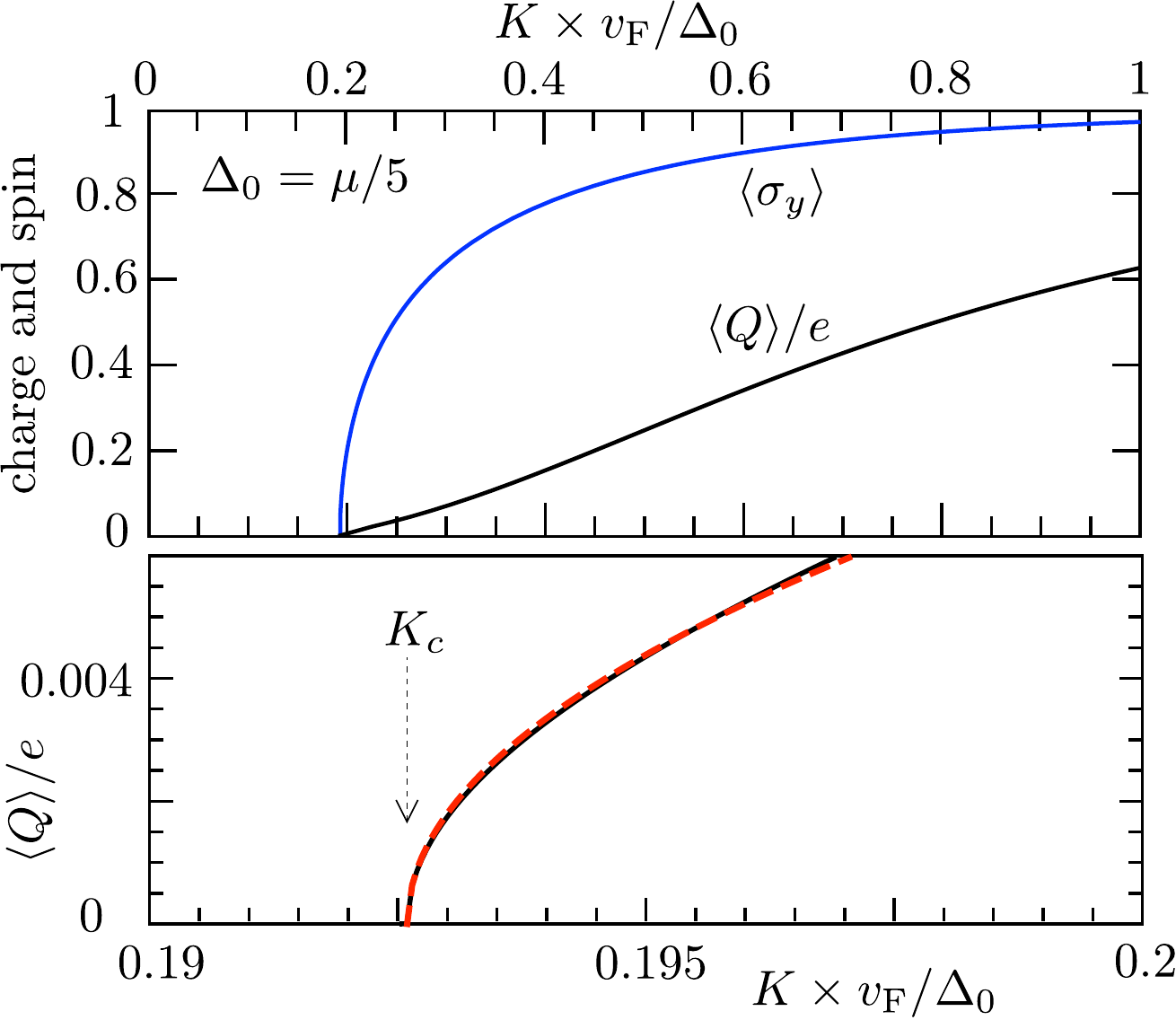}}
\caption{Top panel: Charge and spin expectation value of the Dirac fermions at $E=0$ in the inverted regime. The solid curves are calculated \cite{appendices} in the limit $W\rightarrow\infty$. The lower panel shows a close-up of the charge for $K$ close to $K_c$, the asymptote \eqref{Qapprox} is the dashed curve. The electron-like and hole-like Dirac fermions (red and blue dots in Fig.\ \ref{fig_dispersion1}) have opposite signs of $\langle Q\rangle$ and $\langle\sigma_y\rangle$.
}
\label{fig_charge}
\end{figure}

For $K>K_c$ the inverted Majorana mode at $k_y=0$ coexists with two counterpropagating modes at
\begin{equation}
\pm k_{\rm D}=\pm\sqrt{1+k_{\rm F}/K}\sqrt{K^2+Kk_{\rm F}-(\Delta_0/v_{\rm F})^2}.\label{kDexact}
\end{equation}
Check that $k_{\rm D}=0$ for $K=K_c$. At larger $K$ the Dirac mode momentum $k_{\rm D}$ rises quickly to a value of order $k_{\rm F}$. 

The Dirac fermions have charge expectation value $\pm\langle Q\rangle=\pm e\langle\tau_z\rangle$. Near the transition we find \cite{appendices}
\begin{equation}
\langle Q\rangle= e(\Delta_0/\mu)\sqrt{(K-K_c)/k_{\rm F}},\;\;K\gtrsim K_c.\label{Qapprox}
\end{equation}
As shown in Fig.\ \ref{fig_charge}, the square-root singularity at $K=K_c$ crosses over into an approximately linear increase for larger $K$, up to $Q_{\rm max}=\tfrac{2}{3}e+{\cal O}(\Delta_0/\mu)$ at $K=K^\ast$.

We also show in Fig.\ \ref{fig_charge} that the Dirac mode is approximately spin-polarized, with expectation value $\langle\sigma_y\rangle=\pm(1-\Delta_0/6\mu)$ for $\mu\gg\Delta_0$ and $K$ well above $K_c$. So the Dirac modes differ from the Majorana modes by their nonzero charge and spin expectation value, and there is one more difference: The decay length $\lambda$ of the edge modes into the bulk is smaller for the Dirac modes ($\lambda_{\rm D}\simeq v_{\rm F}/\sqrt{\mu\Delta_0}$) than it is for the Majorana modes ($\lambda_{\rm M}= v_{\rm F}/\Delta_0$, the superconducting coherence length).

\section{No chirality inversion in a \textit{p}-wave superconductor}

The Doppler effect of a supercurrent flowing along the boundary of a spinless chiral \textit{p}-wave superconductor has been studied previously \cite{Yok08,Ser14} --- without producing the chirality inversion we find for the Fu-Kane superconductor. To understand why, we have repeated our calculations for the Hamiltonian
 \begin{equation}
{\cal H}_{\textit{p-wave}}=\begin{pmatrix}
(\bm{k}+\bm{K})^2/2m-\mu&(\Delta_0/k_{\rm F})(k_x+ik_y)\\
(\Delta_0/k_{\rm F})(k_x-ik_y)&\mu-(\bm{k}-\bm{K})^2/2m
\end{pmatrix}\label{Hpwave}
\end{equation}
of a 2D superconductor with a spinless chiral \textit{p}-wave pair potential. Gapless edge modes coexist with a gapped bulk for $\mu=k_{\rm F}^2/2m>0$ and $K<K^\ast=\Delta_0/v_{\rm F}$.

As before, we take a channel of width $W$ along the $y$-axis, parallel to the superflow momentum $\bm{K}=(0,K)$. For large $W$ we find the edge mode dispersion \cite{pwave}
\begin{equation}
E=(v_{\rm F}k_y/k_{\rm F})(K\pm\Delta_0/v_{\rm F}),\label{Ekypwave}
\end{equation}
to first order in $k_y$ and $K$. We see that there is no velocity inversion of the edge modes at any $K<K^\ast$ for which the bulk remains gapped. Comparison with the dispersion \eqref{Ekylargemu} in the Fu-Kane superconductor shows that it is the $\Delta_0$ versus $\Delta_0^2$ dependence that forms the obstruction to $K_c<K^\ast$ in a chiral \textit{p}-wave superconductor.

\begin{figure}[tb]
\centerline{\includegraphics[width=0.7\linewidth]{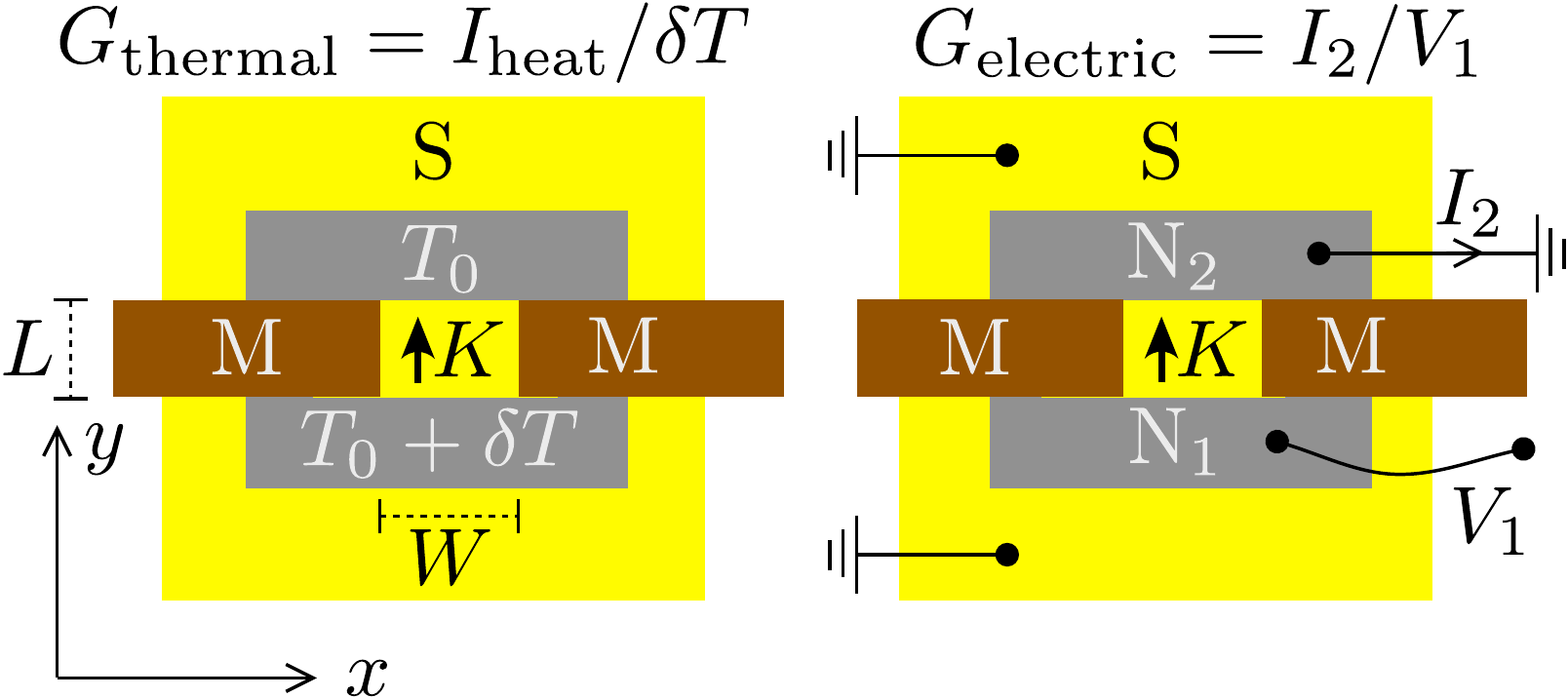}}
\caption{Top view of the proximitized topological insulator ($S$) of Fig.\ \ref{fig_layout}, with additional normal metal contacts ($N_1$, $N_2$) to measure the transport of heat (left panel) and the transport of charge (right panel) through a constriction of width $W$ and length $L$, confined by magnetic insulators ($M$).
}
\label{fig_heatcharge}
\end{figure}

\section{Transport signatures}

The chirality inversion of the edge modes in the Fu-Kane superconductor can be observed in both thermal and electrical conduction. The two transport geometries are shown in Fig.\ \ref{fig_heatcharge}. 

The thermal conductance $G_{\rm thermal}$ at temperature $T_0$ is given by the transmission matrix $t$ (from contact $N_1$ to contact $N_2$),
\begin{equation}
G_{\rm thermal}=G_0\,{\rm Tr}\,t^\dagger t,\;\;G_0=\tfrac{1}{6}(\pi^2 k_{\rm B}^2/h)T_0.\label{eqGthermal}
\end{equation}
The conductance quantum $G_0$ has $1/2$ the value for normal electrons because of the Majorana nature of the carriers. 

The electrical circuit is a three-terminal configuration, with a grounded superconductor in addition to the metal contacts $N_1$, $N_2$. The conductance $G_{\rm electric}=I_2/V_1$, in the zero-temperature, zero-voltage limit, is given by \cite{electric}
\begin{equation}
G_{\rm electric}=(e^2/h)\,{\rm Tr}\,(t_{ee}^\dagger t_{ee}^{\vphantom{\dagger}}-t_{he}^\dagger t_{he}^{\vphantom{\dagger}})=
\frac{e^2}{2h}\,{\rm Tr}\,\tau_z t^\dagger\tau_z t,\label{eqGelectric}
\end{equation}
where $t_{ee}$ and $t_{he}$ are submatrices of $t$ for transmission of an electron as an electron and as a hole, respectively.

\begin{figure}[tb]
\centerline{\includegraphics[width=1\linewidth]{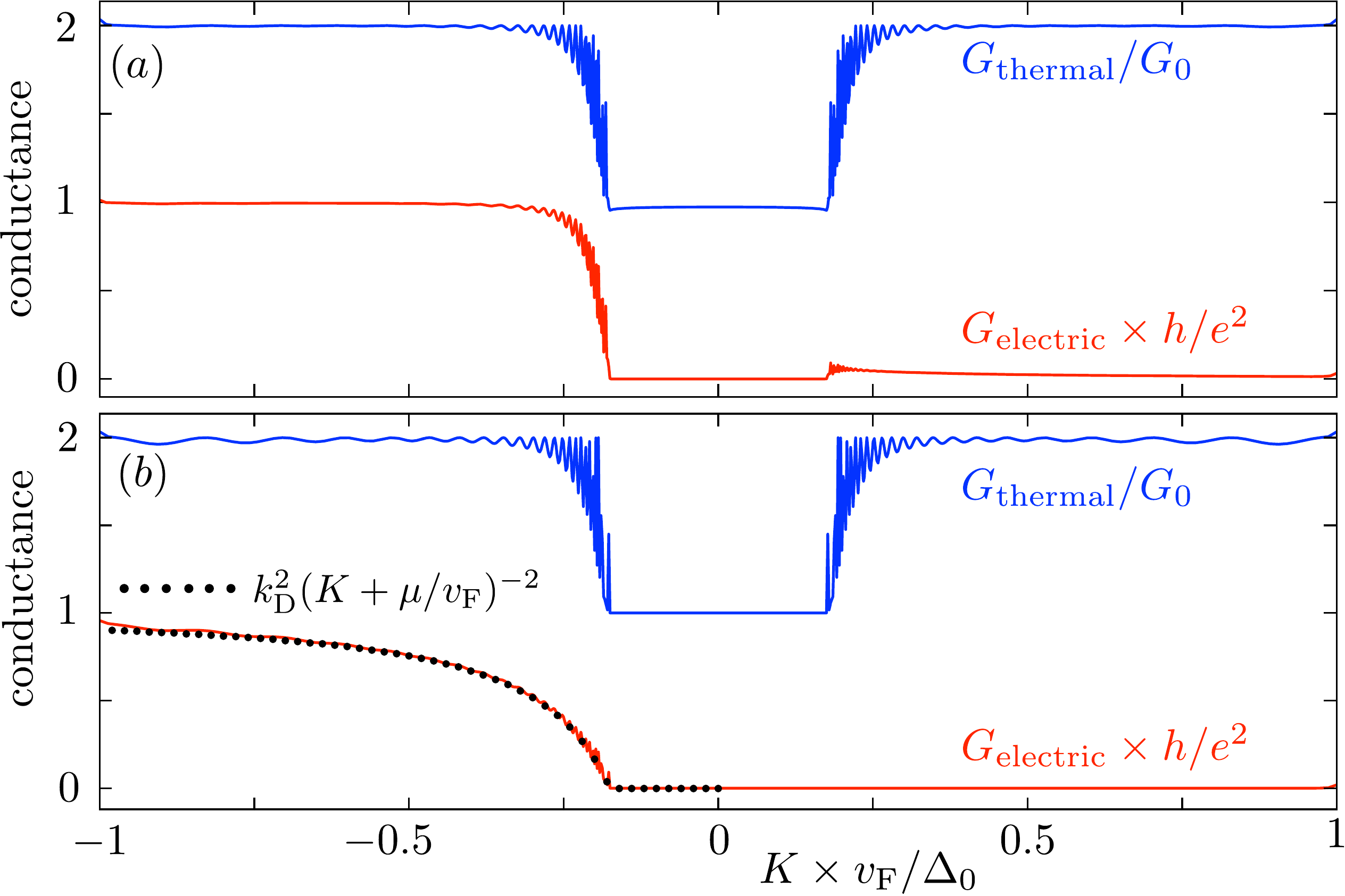}}
\caption{Results for the thermal and electrical conductance, obtained by the numerical simulation \cite{simulation} of a tight-binding model of the Fu-Kane superconductor ($\mu=12\,\Delta_0$, $W=L=1200\,v_{\rm F}/\mu$). In panel (a) the chemical potential $\mu_{\rm N}$ in the normal-metal contacts is equal to the value $\mu$ in the superconducting region, while panel (b) is for the case $\mu_{\rm N}\gg\mu$. The transition starts at a superflow momentum $K$ that is larger than the value $K_c=0.08\,v_{\rm F}/\Delta_0$ from Eq.\ \eqref{Kcexact}, because of the finite lattice spacing ($k_{\rm F}a_0=0.2$). The data points in panel (b) give the analytical result \eqref{Gresult}, with $k_{\rm D}$ from the simulation.
}
\label{fig_transport}
\end{figure}

For $K>K_c$ there are two right-moving edge modes and two left-moving edge modes at the Fermi energy, while for $K<K_c$ there is only a single left-mover and a single right-mover. The thermal conductance is therefore doubled when $K$ becomes larger than $K_c$. For $K<K_c$ the counterpropagating Majorana edge modes carry no electrical charge, while for $K>K_c$ the two co-propagating modes on the same edge form a Dirac mode that can carry charge --- but only in the direction opposite to the superflow. 

To test these expectations we have carried out a numerical simulation of a tight-binding Hamiltonian \cite{simulation}. We compared two models for the normal metal contact, with and without a large potential step at the normal-superconductor (NS) interface. For both models we assumed that the length $L$ of the superconducting region is small compared to the mean free path for disorder scattering, so that any backscattering happens at the NS interfaces. Results are shown in Fig.\ \ref{fig_transport}.

The thermal conductance makes the transition from a completely flat plateau at $G_0$ for $K<K_c$ to a modulated plateau at $2G_0$ for larger $K$.  Because of the appearance of counterpropagating modes on one of the edges the conductance is sensitive to backscattering for $K>K_c$, as is evident by the Fabry-Perot-type oscillations at the onset of the step (when the longitudinal momentum is small). After the onset the plateau is quite flat and close to the quantized value of $2G_0$.

The electrical conductance shows a striking asymmetry in $K$, it remains close to zero for $K>K_c$ and only switches on for $K<-K_c$. This asymmetry under exchance of $N_1$ and $N_2$ is not a violation of reciprocity, since it appears in a three-terminal configuration. The conductance rises to $e^2/h$ in a step-like manner or more slowly, depending on whether or not there is a potential step at the interface.

The reason that the electrical conductance is sensitive to the details of the NS interface, while the thermal conductance is not, is that the heat current from contact ${\rm N}_1$ to contact ${\rm N}_2$ is conserved while the charge current is not. (Charge can be drained into the grounded superconducting terminal, but the gapped superconductor cannot absorb heat.) 

In the absence of a potential step the simulation shows a conductance plateau at $G_{\rm electric}\approx e^2/h$, indicating that a Dirac fermion at $k_{\rm D}\approx k_{\rm F}$ approaching the NS interface transfers a charge $e$ --- notwithstanding its charge expectation value $\langle Q\rangle <e$. We explain this by noting that for $\mu_{\rm N}=\mu$ the longitudinal momentum is approximately conserved across the interface, coupling to states at $-k_{\rm F}$ is suppressed, and since the only outgoing states near $k_{\rm F}$ in the normal region are electrons, the bare charge $e$ is transferred. 

In the presence of a large potential step the longitudinal momentum is not conserved, it is boosted to $+k_{\rm F}$ for the electron component and to $-k_{\rm F}$ for the hole component of the Dirac mode. A mode matching calculation in the limit $\mu_{\rm N}/\mu\rightarrow\infty$ (see App.\ \ref{mode_matching}) gives
\begin{equation}
G_{\rm electric}=\frac{e^2}{h}\frac{k_{\rm D}^2}{(K+\mu/v_{\rm F})^2}=\frac{e^2}{h}\left(1-\frac{(\Delta_0/v_{\rm F})^2}{K(K+\mu/v_{\rm F})}\right),\label{Gresult}
\end{equation}
in excellent agreement with the simulation. 

Eq.\ \eqref{Gresult} can be interpreted in terms of an effective transferred charge, $G_{\rm electric}=(e^\ast)^2/h$ with $e^\ast=ek_{\rm D}/(K+k_{\rm F})$, but $e^\ast$ is very different from $\langle Q\rangle$: While the charge expectation value $\langle Q\rangle$ increases approximately linearly from $0$ to $\frac{2}{3}e$ as $K$ increases from $K_c$ to $K^\ast$ (see Fig.\ \ref{fig_charge}), the effective transferred charge $e^\ast$ increases much more rapidly from $0$ all the way to $e$. We note that in a different charge transfer problem \cite{Lem19}, in a Weyl superconductor, the identification of $e^\ast$ and $\langle Q\rangle$ did hold.

\section{Conclusion}

In summary, we have reported on a manifestation of the Doppler effect from a supercurrent in a spinful topological superconductor: A supercurrent flowing along the magnetic boundary of a Fu-Kane superconductor can reverse the chirality of the Majorana edge mode, without closing the bulk gap. The chirality inversion is accompanied by the appearance of a Dirac mode that propagates counter to the superflow, such that the net number of right-movers minus left-movers is unchanged.

The effect is absent in a spinless chiral \textit{p}-wave superconductor, which is remarkable because the low-energy effective Hamiltonian in the bulk of the Fu-Kane superconductor has $p_x+ip_y$-wave pairing symmetry \cite{Fu08,Yam12}. We have traced the origin of the difference to the linear versus quadratic dependence of the Majorana edge mode velocity on the bulk gap. It is the quadratic dependence that allows the superflow to restructure the edge modes without affecting the bulk spectrum.

The chirality inversion produces a fully electrical signature of the edge currents: charge can be transported upstream relative to the superflow, but not downstream --- because a Majorana mode transports no charge while a Dirac mode does. Such a distinctive effect should help the conclusive observation of chiral Majorana fermions in a topological superconductor.

\acknowledgments

We have benefited from discussions with I. Adagideli. This project has received funding from the Netherlands Organization for Scientific Research (NWO/OCW) and from the European Research Council (ERC) under the European Union's Horizon 2020 research and innovation programme.

\appendix

\section{Calculation of the dispersion relation}
\label{app_dispersion}

The determinantal equation \eqref{detXi}, with the $4\times 4$ matrix $\Omega$ given by Eq.\ \eqref{Omegadef}, is suitable for a numerical calculation of the dispersion relation $E(k_y)$ for finite $W$. Analytical expressions can be obtained in the limit $W\rightarrow\infty$ of uncoupled edges. In this appendix and the next one we set $v_{\rm F}$ to unity, for ease of notation.

\begin{widetext}
The elements of the transfer matrix $e^{W\Omega}$ have an exponential dependence $\propto e^{W\xi_\pm}$ and $\propto e^{-W\xi_\pm}$ on $W$, with
\begin{equation}
\xi_\pm= \sqrt{\Delta_0^2-E^2-\mu^2+k_y^2+K^2\pm 2 \sqrt{\Delta_0^2 \left(K^2-\mu^2\right)+(k_y K-E \mu)^2}}.
\end{equation} 
The sign ambiguity in the square roots is resolved by taking the square root with a positive real part (branch cut along the negative real axis). The edge modes in the limit $W\rightarrow\infty$ are obtained by setting $e^{-W\xi_\pm}\rightarrow 0$ in the transfer matrix. The determinantal equation \eqref{detXi} then reduces to
\begin{equation}
\Delta_0^2(K^2-\mu^2)(E^2-k_y^2-K^2+\mu^2)+(\Delta_0^2-\xi_-\xi_+) \left(\Delta_0^2(K^2-\mu^2)+2 (k_y K-E \mu)^2\right)=0.\label{determinant_step1}
\end{equation}
We eliminate the square roots in the product $\xi_-\xi_+$ by rearranging the equation as $\xi_-\xi_+=\cdots$ and then squaring both sides, resulting in
\begin{align}
\bigl(\Delta_0^2 -(E-K)^2+(k_y-\mu )^2\bigr) \bigl(\Delta_0^2-(E+K)^2 +(k_y+\mu )^2\bigr)={}&\Delta_0^4\left(1+\frac{(K^2-\mu^2)(E^2-k_y^2-K^2+\mu^2)}{\Delta_0^2 (K^2-\mu^2)+2 (k_yK-E\mu)^2}\right)^2.\label{determinant_step2}
\end{align}
\end{widetext}

Eq.\ \eqref{determinant_step2} has eight solutions for $E$, the two physical solutions are the dispersions $E_\pm(k_y)$ that cross zero at $k_y=0$. The full expressions are a bit lengthy and not recorded here. The linear dispersion near $k_y=0$ does have a compact expression, given by Eq.\ \eqref{Ekylargemu} in the main text. Eq.\ \eqref{Kcexact} for $K_c$ is the value of $K$ at which the slope of $E_-(k_y)$ vanishes. To find the momenta $k_y=\pm k_{\rm D}$ of the Dirac modes for $K>K_c$ we solve Eq.\ \eqref{determinant_step2} for $k_y$ at $E=0$, resulting in Eq.\ \eqref{kDexact}.

\section{Calculation of the charge and spin of the Dirac mode}
\label{app_charge}

The charge expectation value $\langle Q\rangle=e\langle\tau_z\rangle$ can be obtained from the dispersion relation via the derivative $\langle Q\rangle=-e\partial E/\partial \mu$. It vanishes for the Majorana fermions at $k_y=0$, but it is nonzero for the Dirac fermions at $k_y=\pm k_{\rm D}$, with $E(k_{\rm D})=0$. We can compute this directly from the determinantal equation \eqref{determinant_step2}, by substituting $E\mapsto E(\mu)$, differentiating with respect to $\mu$, solving for $E'(\mu)$, and finally setting $E(\mu)\mapsto 0$, $k_y\mapsto k_{\rm D}$.

We thus arrive at the Dirac fermion charge
\begin{equation}
\langle Q\rangle=e\frac{\sqrt{K} \sqrt{K (K+\mu )-\Delta_0^2} \left(2 K (K+\mu )-\Delta_0^2\right)}{\sqrt{K+\mu } \left[\Delta_0^2 (\mu -K)+2 K^2 (K+\mu )\right]}.\label{chargeformula}
\end{equation}
This is the black curve plotted in the top panel of Fig.\ \ref{fig_charge}. Expansion near $K=K_c$ gives for $\mu\gg \Delta_0$ the square-root result \eqref{Qapprox} in the main text. The charge increases monotonically with increasing $K$, reaching its maximal value
\begin{equation}
Q_{\rm max}=e\sqrt{\frac{\mu}{\Delta_0+\mu}}\;\frac{\Delta_0+2\mu}{\Delta_0+3\mu}\label{Qmax}
\end{equation}
at $K=K^\ast=\Delta_0$.

In a similar way we can calculate the spin expectation value $\langle \sigma_y\rangle=\partial E/\partial K$ of the Dirac fermions, with the result
\begin{equation}
\langle\sigma_y\rangle=\frac{\sqrt{K (K+\mu )-\Delta_0^2} \left(2 K^2 (K+\mu )+\Delta_0^2 \mu \right)}{ \sqrt{K(K+\mu) } \left[\Delta_0^2 (\mu -K)+2 K^2 (K+\mu )\right]},\label{spinformula}
\end{equation}
see the blue curve in Fig.\ \ref{fig_charge}. The behavior for $K\gtrsim K_c$ is again a square root increase, $\langle\sigma_y\rangle\approx (\sqrt\mu/\Delta_0)\sqrt{K-K_c}$, rising rapidly to a value 
\begin{equation}
\langle\sigma_y\rangle_{\rm max}=\sqrt{\frac{\mu}{\Delta_0+\mu}}\,\frac{2\Delta_0+3\mu}{\Delta_0+3\mu}\rightarrow 1-\frac{\Delta_0}{6\mu}\;\;\text{for}\;\;\mu\gg\Delta_0,
\end{equation}
close to unity.

The signs of spin and charge are such that $\langle Q\rangle<0$ and $\langle\sigma_y\rangle>0$ for the Dirac mode at $k_y=k_{\rm D}$. The mode at $k_y=-k_{\rm D}$ has the opposite signs.

\section{Doppler-boosted edge modes in a chiral \textit{p}-wave superconductor}
\label{app_pwave}

The chiral \textit{p}-wave Hamiltonian has the form
\begin{subequations}
\begin{align}
&{\cal H}_{\text{p-wave}}=\begin{pmatrix}
H_0&\hat{\Delta}\\
\hat{\Delta}^\dagger&-H_0^\ast
\end{pmatrix},\\
&H_0=\frac{1}{2m}(k_x^2+k_y^2)-\mu,\\
&\hat{\Delta}=k_{\rm F}^{-1}\{\Delta(\bm{r}),k_x+ik_y\},
\end{align}
\end{subequations}
with $\bm{k}=-i\partial/\partial\bm{r}$ and $\{a,b\}=\tfrac{1}{2}(ab+ba)$ the symmetrization operator. 

The superflow momentum $\bm{K}$ enters in the pair potential via $\Delta(\bm{r})=\Delta_0 e^{2i\bm{K}\cdot\bm{r}}$. We remove it by a gauge transformation,
\begin{equation}
{\cal H}_{\text{p-wave}}\mapsto U^\dagger H_{\text{p-wave}}U,\;\;U=
\begin{pmatrix}
e^{i\bm{K}\cdot\bm{r}}&0\\
0&e^{-i\bm{K}\cdot\bm{r}}\end{pmatrix}.
\end{equation}
In view of the identity
\begin{equation}
e^{-i\bm{K}\cdot\bm{r}}\{e^{i\bm{K}\cdot\bm{r}},\partial_x+i\partial_y\}e^{-i\bm{K}\cdot\bm{r}}=\partial_x+i\partial_y,
\end{equation}
the transformed Hamiltonian \eqref{Hpwave} contains the Doppler shifted momentum only in the diagonal elements, not in the off-diagonal elements. 

In terms of the Pauli matrices $\tau_\alpha$ acting on the electron-hole degree of freedom, we have
\begin{equation}
{\cal H}_{\text{p-wave}}=\frac{k^2}{2m}\tau_z-\mu\tau_z+\frac{K}{m}k_y\tau_0+\frac{\Delta_0}{k_{\rm F}}(k_x\tau_x-k_y\tau_y),
\end{equation}
to first order in $\bm{K}=(0,K)$. We introduce a boundary at $x=0$ and seek the velocity of an edge mode in the $y$-direction. The velocity operator at $k_y=0$ is
\begin{equation}
\hat{v}_{\rm edge}=\lim_{k_y\rightarrow 0}\frac{\partial}{\partial k_y}{\cal H}_{\text{p-wave}}=\frac{K}{m}\tau_0-\frac{\Delta_0}{k_{\rm F}}\tau_y.
\end{equation}

The edge mode wave function at $E=0$, $k_y=0$ solves
\begin{equation}
\frac{1}{2m}\psi''(x)+\mu\psi(x)-\frac{\Delta_0}{k_{\rm F}}\tau_y\psi'(x)=0,
\end{equation}
for $x>0$, with boundary condition $\psi(0)=0$. A normalizable solution exists for $\mu>0$, it is an eigenstate of $\tau_y$ with eigenvalue $-1$. The expectation value $v_{\rm edge}$ of the velocity follows directly,
\begin{equation}
v_{\rm edge}=\langle\psi|\hat{v}_{\rm edge}|\psi\rangle=K/m+\Delta_0/k_{\rm F}.
\end{equation}
At the opposite edge the solution $\psi$ is an eigenstate of $\tau_y$ with eigenvalue $+1$, resulting in a velocity $v_{\rm edge}=K/m-\Delta_0/k_{\rm F}$. The corresponding edge mode dispersion is given by Eq.\ \eqref{Ekypwave}.

\section{Details of the tight-binding simulation}
\label{app_simulation}

For the numerical calculations we model the Fu-Kane superconductor by a tight-binding Hamiltonian on a 2D square lattice (lattice constant $a_0$),
\begin{align}
{\cal H}={}&\frac{v_{\rm F}}{a_0}\sum_{\alpha=x,y}\sin(a_0 k_\alpha+a_0 K_\alpha)\sigma_\alpha\tau_z\nonumber\\
&+M\sigma_z\tau_0-\mu\sigma_0  \tau_z+\Delta_0\sigma_0 \tau_x\nonumber\\
&+\frac{M_0 v_{\rm F}}{a_0}\sum_{\alpha=x,y}[1-\cos(a_0k_\alpha+a_0K_\alpha)]\sigma_z \tau_0.\label{HS}
\end{align}
In the limit $a_0\rightarrow 0$ the continuum Hamiltonian \eqref{Hfourband} is recovered. The term $\propto M_0$ is introduced to avoid spurious Dirac points at the edge of the Brillouin zone (fermion doubling).

We consider a channel geometry of width $W$ along the $y$-axis, with mass $M=0$ for $|x|\leq W/2$ and infinite mass $M\rightarrow\infty$ for $|x|>W/2$. It is efficient if we can replace the infinite-mass term by a lattice termination at $|x|=W/2$, so that we only have to consider the lattice points inside the channel. This is allowed if the lattice termination enforces the boundary condition \eqref{boundcond}. We can set $M_0=-1$ to achieve that goal.

To see this, consider the matrix elements for hopping in the $\pm x$-direction,
\begin{equation}
{\cal H}_{n_{x}\pm 1,n_x}=\pm \frac{v_{\rm F}}{2ia_0}e^{\pm ia_0K_x}\sigma_x\tau_z-\frac{M_0v_{\rm F}}{2a_0}e^{\pm ia_0K_x}\sigma_z\tau_0.
\end{equation}
To represent the boundary condition \eqref{boundcond} by a lattice termination at the right edge, we need to ensure that ${\cal H}_{n_{x}- 1,n_x}\psi=0$ at $x=W/2+a_0$ when $\psi=+\sigma_y\tau_z\psi$. Similarly, for $x=-W/2$ we need ${\cal H}_{n_{x}+1,n_x}\psi=0$ at $x=-W/2-a_0$ when $\psi=-\sigma_y\tau_z\psi$. One readily checks that both conditions are realized if $M_0=-1$.

\begin{figure}[tb]
\centerline{\includegraphics[width=0.8\linewidth]{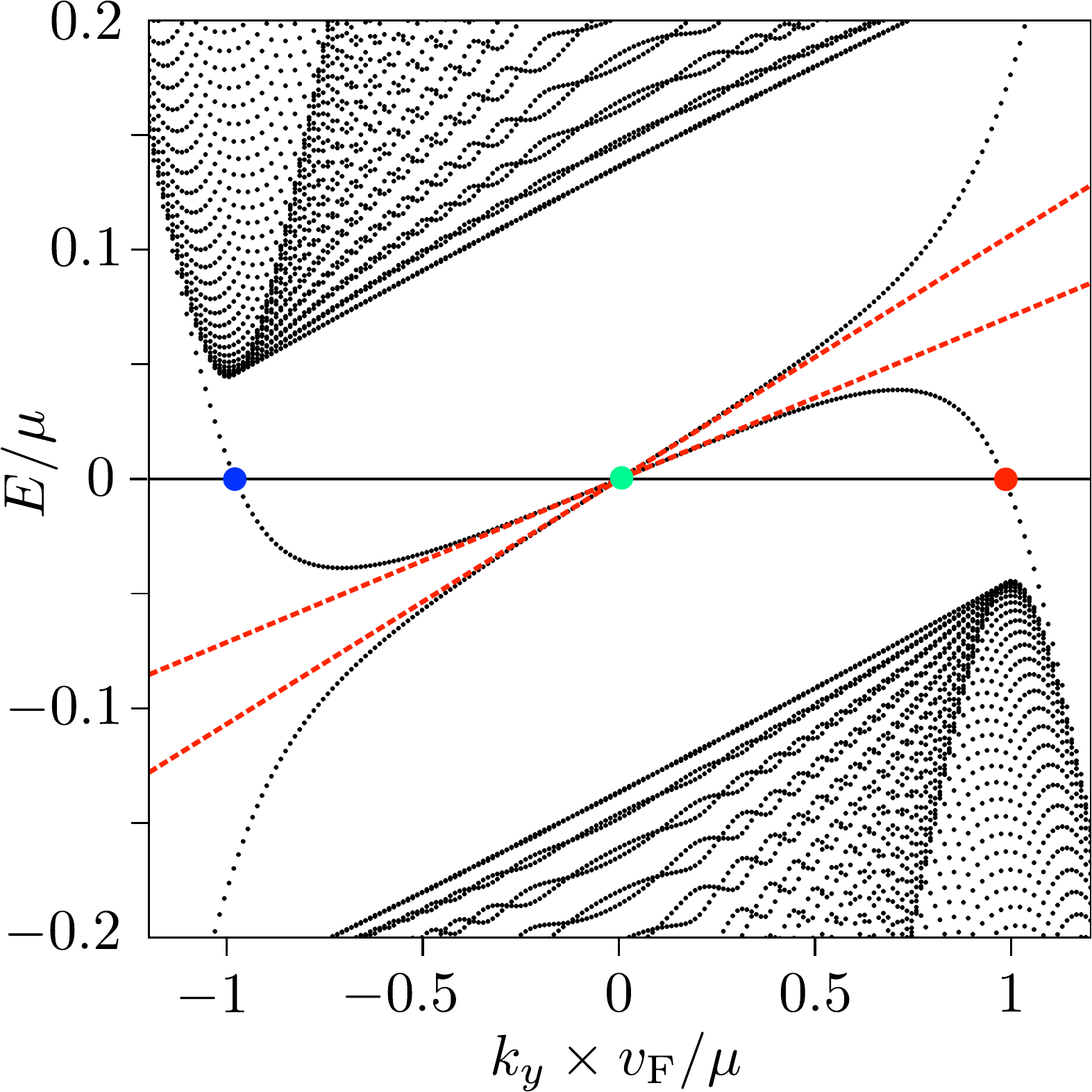}}
\caption{Energy spectrum of the superconducting channel, calculated numerically from the tight-binding Hamiltonian \eqref{HS} ($\mu=7.5\,\Delta_0$, $W=100\,v_{\rm F}/\mu$, $K=\tfrac{2}{3}\Delta_0/v_{\rm F}$, $a_0=0.02\,v_{\rm F}/\mu$). The red dashed lines are the large-$\mu$, large-$W$ asymptotes \eqref{Ekylargemu} of the Majorana edge mode dispersion. The red and blue dots indicate the Dirac fermion mode at $k_y=\pm k_{\rm D}$, the green dot is the Majorana fermion at $k_y=0$.
}
\label{fig_simulationboundary}
\end{figure}

\begin{figure}[tb]
\centerline{\includegraphics[width=0.8\linewidth]{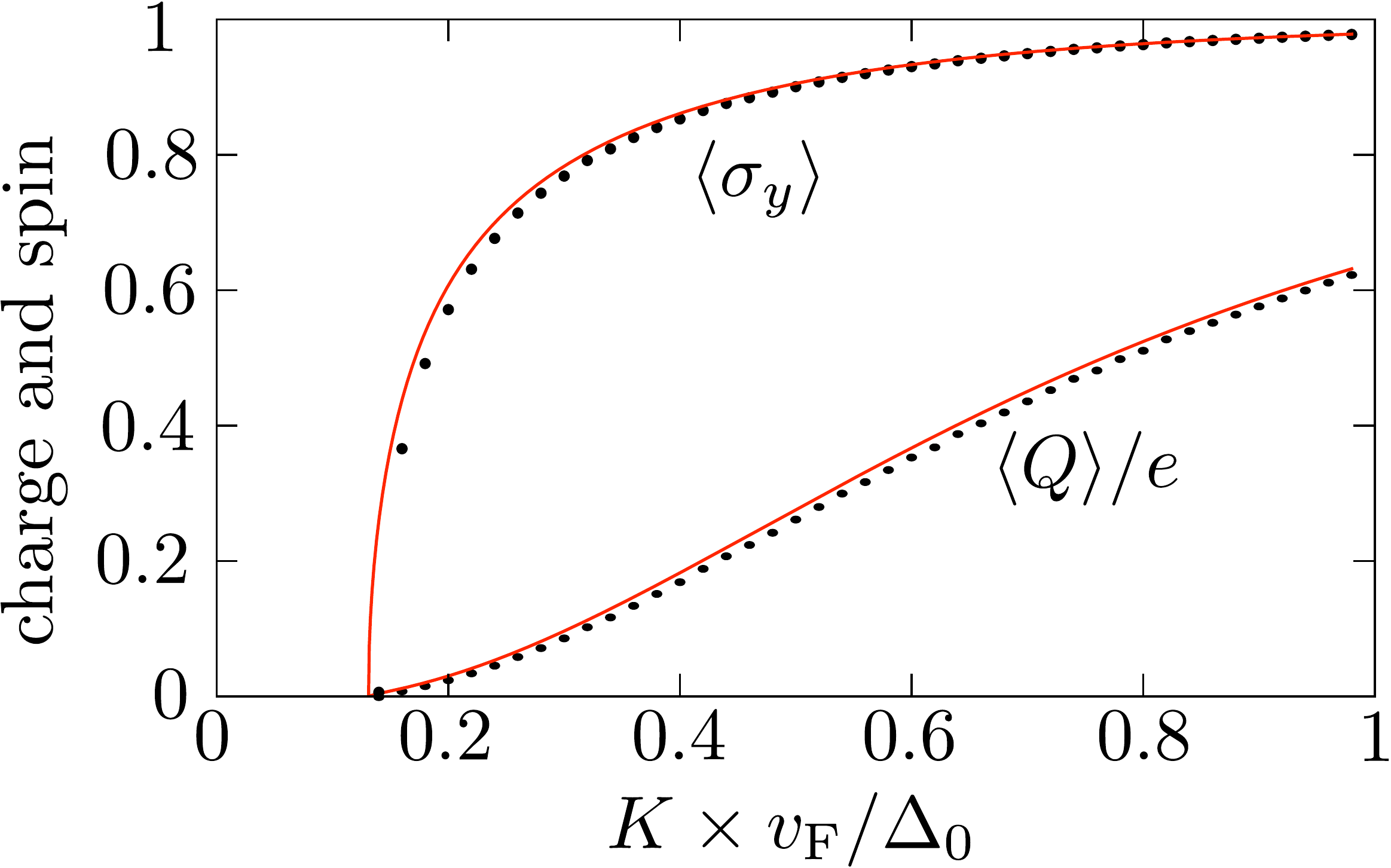}}
\caption{Expectation value of the charge and the spin of the Dirac fermions in the inverted regime, as a function of the superflow momentum $K$. The data points result from the tight-binding simulation (same parameters as in Fig.\ \ref{fig_simulationboundary}), the solid curves are the analytical results \eqref{chargeformula} and \eqref{spinformula} in the limit $W\rightarrow\infty$.
}
\label{fig_simulationcharge}
\end{figure}

In Figs.\ \ref{fig_simulationboundary} and \ref{fig_simulationcharge} we show that we recover the analytical results for the edge mode dispersion and for the expectation value of the charge and spin of the Dirac fermions. To achieve this accurate agreement the tight-binding model needs to be close to the continuum limit. For that purpose we took a small lattice contant ($k_{\rm F}a_0=0.02$), which is computationally feasible in an effectively 1D simulation. The transport calculations are fully 2D and we were forced to take a ten times larger lattice constant to keep the problem tractable. This is why the numerical value of $K_c$ in Fig.\ \ref{fig_transport} differs substantially from the analytical result in the continuum limit.

\begin{figure}[tb]
\centerline{\includegraphics[width=0.6\linewidth]{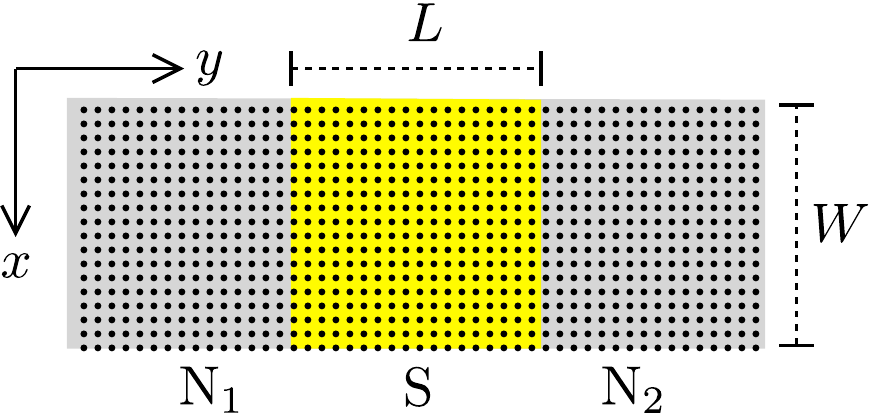}}
\caption{Two-dimensional square lattice on which the tight-binding model is defined. The Hamiltonian \eqref{HS} is applied to the superconducting segment of length $L$ (yellow). In the semi-infinite leads (grey) the Hamiltonian is given by Eqs.\ \eqref{Hlead2} and \eqref{Hlead}, respectively, in the models with and without a potential step at the NS interfaces.
}
\label{fig_layouttb}
\end{figure}

For the transport calculations we take a finite length $L$ of the superconducting segment (S), and attach semi-infinite normal metal leads (N) at the two ends (see Fig.\ \ref{fig_layouttb}). We set $\Delta_0=0$ in N, no coupling of electrons and holes (the value of $K$ then becomes irrelevant and may be set to zero),
\begin{align}
{\cal H}_{\rm lead}={}&\frac{v_{\rm F}}{a_0}\sum_{\alpha=x,y}\sigma_\alpha\tau_z\sin a_0 k_\alpha -\mu\sigma_0  \tau_z \nonumber\\
&+\frac{M_0v_{\rm F}}{a_0}\sum_{\alpha=x,y}(1-\cos a_0k_\alpha)\sigma_z \tau_0.\label{Hlead}
\end{align}
We again set $M_0=-1$ to implement the infinite-mass boundary condition by a lattice termination at $x=\pm W/2$.

Eq.\ \eqref{Hlead} is the model without a potential step at the NS interface (panel \textit{a} in Fig.\ \ref{fig_transport}). If the chemical potential $\mu_{\rm N}$ in the normal metal leads is much larger than the value $\mu$ in the superconducting region, only modes with a large longitudinal momentum $k_y$ are transmitted across the NS interface. We cannot directly take the large-$\mu_{\rm N}$ limit in the Hamiltonian \eqref{Hlead}, because of the finite band width. Instead, we achieve the same goal of suppressing transverse momenta by cutting the transverse hoppings at $\mu_{\rm N}=0$,
\begin{align}
{\cal H}_{\rm lead}(\text{large potential step})=\frac{v_{\rm F}}{a_0}\sigma_y\tau_z\sin a_0 k_y \nonumber\\
{}+\frac{M_0v_{\rm F}}{a_0}(1-\cos a_0k_y)\sigma_z \tau_0.\label{Hlead2}
\end{align}
This produces the data in panel \textit{b} of Fig.\ \ref{fig_transport}.

\section{Derivation of Eq.\ (\ref{Gresult})} 
\label{mode_matching}

\subsection{Calculation of the transferred charge}

We seek to compute the charge $e^\ast$ transferred across the NS interface at $y=0$ by a Dirac fermion at $k_y=\pm k_{\rm D}$. We assume a large potential step at the interface, such that the chemical potential $\mu_{\rm N}$ in the normal region $y<0$ is much larger than the value $\mu=v_{\rm F}k_{\rm F}$ in the superconducting region $y>0$. The Hamiltonian in S is
\begin{align}
{\cal H}={}&v_{\rm F}(k_x\sigma_x+k_y\sigma_y)\tau_z+v_{\rm F}K\sigma_y\tau_0-\mu\sigma_0\tau_z\nonumber\\
&+M\sigma_z\tau_0+\Delta_0\sigma_0\tau_x\nonumber\\
\equiv{}&H_0+v_{\rm F}k_y\sigma_y\tau_z.
\end{align}
For later use we have separated out the $k_y$-independent part $H_0=\lim_{k_y\rightarrow 0}{\cal H}$. 

The potential step boosts the momentum component $k_y$ perpendicular to the interface, without affecting the parallel component $k_x$, so in N only modes are excited with $|k_y|\gg |k_x|$. These are eigenstates of $\sigma_y\tau_z$ with eigenvalue $-1$, moving away from the interface in the $-y$ direction. Continuity of the wave function $\Psi$ at the interface then requires that $\lim_{y\rightarrow 0}\Psi\equiv\Psi_0$ satisfies
\begin{equation}
\sigma_y\tau_z\Psi_0=-\Psi_0\Leftrightarrow {\cal P}\Psi_0=\Psi_0,
\end{equation}
with projection operator
\begin{equation}
{\cal P}=\tfrac{1}{2}(1-\sigma_y\tau_z).
\end{equation}

The eigenvalue equation ${\cal H}\Psi=0$ at $E=0$ implies that
\begin{align}
0&=\lim_{y\downarrow 0}{\cal P}{\cal H}\Psi={\cal P}H_0{\cal P}\Psi_0+\lim_{y\downarrow 0}{\cal P}v_{\rm F}k_y\sigma_y\tau_z\Psi\nonumber\\
&=(K+k_{\rm F}){\cal P}\hat{j}{\cal P}\Psi_0-i{\cal P}\hat{v}\Psi'_0,\label{Kplusmu}
\end{align}
with the definitions $\hat{j}=v_{\rm F}\sigma_y\tau_0$, $\hat{v}=v_{\rm F}\sigma_y\tau_z$, and $\Psi'_0=\lim_{y\downarrow 0}\partial\Psi/\partial y$. The derivative is not continuous at the NS interface, hence the specification that the limit $y\downarrow 0$ should be taken from above. Also note that ${\cal P}\Psi_0=\Psi_0$ but ${\cal P}\Psi'_0\neq\Psi'_0$.

We define the $y$-dependent inner product of two arbitrary states,
\begin{equation}
\langle\Psi_1|\Psi_2\rangle_y=\int dx\, \Psi_1^\ast(x,y)\Psi_2(x,y).
\end{equation}
With respect to this inner product the operator $H_0$ is self-conjugate, $\langle\Psi_1|H_0\Psi_2\rangle_y=\langle H_0\Psi_1|\Psi_2\rangle_y$, but the operator $k_y=-i\partial/\partial y$ is not (an integration over $y$ would be needed for that). Still, if $\Psi$ is an eigenstate of ${\cal H}$ at eigenvalue $E$, we have $k_y \hat{v}\Psi=(E-H_0)\Psi$, so $k_y\hat{v}$ inherits the self-conjugate property from $H_0$, $\langle\Psi|k_y\hat{v}\Psi\rangle_y=\langle k_y\hat{v}\Psi|\Psi\rangle_y$. 

We will use this identity in the two forms
\begin{equation}
\langle\Psi|\hat{v}|\Psi'\rangle_y=-\langle\Psi'|\hat{v}|\Psi\rangle_y,\;\;\langle\Psi'|\hat{v}|\Psi'\rangle_y=-\langle\Psi|\hat{v}|\Psi''\rangle_y,\label{PsiPsiprimeidentity}
\end{equation}
where $\Psi'=\partial\Psi/\partial y$ and $\Psi''=\partial^2\Psi/\partial y^2$. (The second equality holds because ${\cal H}$ does not depend on $y$, so if ${\cal H}\Psi=E\Psi$ then also ${\cal H}\Psi'=E\Psi'$.)

One implication of Eq.\ \eqref{PsiPsiprimeidentity} is that the particle current $\langle\Psi|\hat{v}|\Psi\rangle_y$ is $y$-independent, as it should be,
\begin{equation}
\frac{d}{dy}\langle\Psi|\hat{v}|\Psi\rangle_y=\langle\Psi'|\hat{v}|\Psi\rangle_y+\langle\Psi|\hat{v}|\Psi'\rangle_y=0.\label{conservation1}
\end{equation}
A more unexpected implication is that also the expectation value $\langle\Psi|\hat{v}|\Psi'\rangle_y$ is $y$-independent,
\begin{equation}
\frac{d}{dy}\langle\Psi|\hat{v}|\Psi'\rangle_y=\langle\Psi'|\hat{v}|\Psi'\rangle_y+\langle\Psi|\hat{v}|\Psi''\rangle_y=0.\label{conservation2}
\end{equation}
We will make essential use of these two properties in just a moment.

The charge current $I_{\rm charge}$ through the NS interface at $y=0$,
\begin{equation}
I_{\rm charge}=e \langle\Psi|\hat{j}|\Psi\rangle_0=e\langle \Psi|{\cal P}\hat{j}{\cal P}|\Psi\rangle_0,
\end{equation}
can be rewritten by substitution of Eq.\ \eqref{Kplusmu},
\begin{equation}
I_{\rm charge}=\frac{ie}{K+k_{\rm F}}\langle\Psi|{\cal P}\hat{v}|\Psi'\rangle_0=\frac{ie}{K+k_{\rm F}}\langle\Psi|\hat{v}|\Psi'\rangle_0.\label{Icharge}
\end{equation}
The renormalized charge $e^\ast$ transferred through the NS interface by a Dirac fermion is the ratio of the charge current and the particle current $I_{\rm particle}=\langle\Psi|\hat{v}|\Psi\rangle_0$,
\begin{equation}
e^\ast=\frac{ie}{K+k_{\rm F}}\frac{\langle\Psi|\hat{v}|\Psi'\rangle_0}{\langle\Psi|\hat{v}|\Psi\rangle_0}=\frac{ie}{K+k_{\rm F}}\frac{\langle\Psi|\hat{v}|\Psi'\rangle_y}{\langle\Psi|\hat{v}|\Psi\rangle_y}.\label{eastintermediate}
\end{equation}
In the second equality we used Eqs.\ \eqref{conservation1} and \eqref{conservation2}.

We can evaluate the ratio of $y$-dependent expectation values at large $y$, far from the interface, where evanescent waves have decayed to zero and $\Psi$ contains only the propagating Dirac mode $\Psi_{\rm D}\propto e^{\pm ik_{\rm D}y}$ --- under the assumption that there is no backscattering of quasiparticles at the interface. The ratio then reduces to $\pm ik_{\rm D}$, resulting in a transferred charge
\begin{equation}
\pm e^\ast=\pm \frac{ek_{\rm D}}{K+k_{\rm F}}.\label{estarresult}
\end{equation}
The sign of the transferred charge is set by the sign of the charge expectation value $\langle Q\rangle $ of the Dirac mode, but the magnitude is different.

Eq.\ \eqref{estarresult} gives the charge of an outgoing mode in N (moving away from the NS interface), when it is matched to an incoming Dirac mode in S (moving towards the NS interface). The entire calculation carries over if the direction of motion is inverted, so when an incoming mode in N is matched to an outgoing Dirac mode in S, the incoming mode has the same charge $\pm e^\ast$.

\subsection{Calculation of the electrical conductance}

The transferred charge determines the conductance $G_{\rm electric}=I_2/V_1$ that gives the electrical current $I_2$ into the normal contact ${\rm N}_2$ in response to a voltage $V_1$ applied to contact ${\rm N}_1$ (see Fig.\ \ref{fig_layouttb}). This is a three-terminal circuit, the third terminal is the grounded superconductor S connecting ${\rm N}_1$ and ${\rm N}_2$, separated by a distance $L$. We assume that both contacts have a chemical potential $\mu_{\rm N}\gg \mu$.

In the absence of backscattering the transmission matrix $t$ from ${\rm N}_1$ to ${\rm N}_2$ is a rank-two matrix of the form
\begin{equation}
t=e^{ik_{\rm D}L}|\Psi^+_2\rangle\langle\Psi^+_1|+e^{-ik_{\rm D}L}|\Psi^-_2\rangle\langle\Psi^-_1|.\label{tsecondrank}
\end{equation}
The incoming mode $|\Psi^\pm_1\rangle$ in contact ${\rm N}_1$ is matched in S to a Dirac mode at $k_y=\pm k_{\rm D}$. The Dirac mode propagates to contact ${\rm N}_2$, picking up a phase $e^{\pm i k_{\rm D}L}$, and is then matched to an outgoing mode $|\Psi^\pm_2\rangle$. The  matching condition gives a charge $\pm e^\ast$ to $\Psi^\pm_n$,
\begin{equation}
\langle\Psi^\pm_n|\tau_z|\Psi^\pm_n\rangle=\pm e^\ast.\label{Psipmeast}
\end{equation}

The modes $|\Psi^+_n\rangle$ and $|\Psi^-_n\rangle$ not only carry opposite charge, they are each others particle-hole conjugate,
\begin{equation}
|\Psi^+_n\rangle=\sigma_y\tau_y|\Psi^-_n\rangle^\ast,
\end{equation}
as they are matched to Dirac modes that are related by particle-hole conjugation. We will use an orthogonality consequence of this property:
\begin{align}
\langle\Psi^+_n|\tau_z|\Psi^-_n\rangle&=-\langle\Psi^+_n|\sigma_y\tau_y\tau_z\sigma_y\tau_y|\Psi^-_n\rangle=-\langle\Psi^-_n|\tau_z|\Psi^+_n\rangle^\ast\nonumber\\
&=-\langle\Psi^+_n|\tau_z|\Psi^-_n\rangle\Rightarrow\langle\Psi^+_n|\tau_z|\Psi^-_n\rangle=0.\label{orthogonality}
\end{align}
So while current conservation by itself requires that $|\Psi^+_n\rangle$ is orthogonal to $|\Psi^-_n\rangle$, the additional constraint of particle-hole symmetry also gives the orthogonality of $|\Psi^+_n\rangle$ and $\tau_z|\Psi^-_n\rangle$.

We now have all the pieces in place to calculate the conductance, given in terms of the transmission matrix by
\begin{equation}
G_{\rm electric}=\frac{e^2}{2h}\,{\rm Tr}\,\tau_z t^\dagger \tau_z t.
\end{equation}
Substitution of Eq.\ \eqref{tsecondrank} and use of the orthogonality \eqref{orthogonality} gives
\begin{equation}
G_{\rm electric}=\frac{e^2}{2h}\sum_{s=\pm}\,\langle\Psi^s_2|\tau_z|\Psi^s_2\rangle\langle\Psi^s_1|\tau_z|\Psi^s_1\rangle=\frac{(e^\ast)^2}{h},
\end{equation}
where in the second equality we used Eq.\ \eqref{Psipmeast}. Subsitution of Eq.\ \eqref{estarresult} then produces Eq.\ \eqref{Gresult} in the main text.


\begin{thebibliography}{99}
\bibitem{Rashba} The Hamiltonian \eqref{Hfourband} locks the spin in the direction of motion, via the term $k_x\sigma_x+k_y\sigma_y$. Rashba spin-orbit coupling would lock the spin in the perpendicular direction, via $k_x\sigma_y-k_y\sigma_x$. The two Hamiltonians are related by a unitary transformation, so we can choose one type of coupling without loss of generality. The $\sigma_y$-polarization of the Dirac edge mode that we find for the Hamiltonian \eqref{Hfourband} (with a boundary parallel to the $y$-axis) corresponds to a $\sigma_x$-polarization for the Rashba Hamiltonian.
\bibitem{appendices} Details of the calculation of the edge mode dispersion are given in App.\ \ref{app_dispersion}; details of the calculation of the charge and spin of the Dirac mode are given in App.\ \ref{app_charge}.
\bibitem{pwave} That the shift of $\bm{k}$ by the superflow momentum $\bm{K}$ appears only in the diagonal elements of the Hamiltonian \eqref{Hpwave} is required by gauge invariance, see App.\ \ref{app_pwave}, which also contains a derivation of Eq.\ \eqref{Ekypwave}.
\bibitem{electric} The second equality in Eq.\ \eqref{eqGelectric} follows from the first by rewriting $G_{\rm electric}/G_0=(1/2)\,{\rm Tr}\,(1+\tau_z)t^\dagger\tau_z t$, and then using particle-hole symmetry to identify ${\rm Tr}\,t^\dagger\tau_z t=0$.
\end{thebibliography}
\end{document}